\documentclass[aps,prb,reprint,preprintnumbers,amsmath,amssymb,superscriptaddress, showpacs,showkeys,floatfix]{revtex4-1}

\usepackage{graphicx}
\usepackage{dcolumn}
\usepackage{bm}
\usepackage{color} 
\usepackage{braket}
\usepackage[separate-uncertainty,multi-part-units = single ]{siunitx}
\usepackage{lipsum}
\usepackage{float}
\usepackage{bm}

\usepackage[english]{babel}
\usepackage[
	colorlinks=true,
	frenchlinks=false,
        linkcolor=blue,
	    anchorcolor=blue,
        citecolor=blue,
        filecolor=blue,
        urlcolor=blue,
        bookmarks=true,
        bookmarksopen=true,
	    bookmarksnumbered=true, 
        bookmarksopenlevel=1,
        plainpages=false,
        pdfpagelabels=true,
	    draft=false]{hyperref}

\begin{document}

\title[]{Secure quantum remote state preparation of squeezed microwave states}

\date{\today}

\author{S.~Pogorzalek}
\email[]{stefan.pogorzalek@wmi.badw.de}
\affiliation{Walther-Mei{\ss}ner-Institut, Bayerische Akademie der Wissenschaften, 85748 Garching, Germany}
\affiliation{Physik-Department, Technische Universit\"{a}t M\"{u}nchen, 85748 Garching, Germany}

\author{K. G. Fedorov}
\email[]{kirill.fedorov@wmi.badw.de}
\affiliation{Walther-Mei{\ss}ner-Institut, Bayerische Akademie der Wissenschaften, 85748 Garching, Germany}
\affiliation{Physik-Department, Technische Universit\"{a}t M\"{u}nchen, 85748 Garching, Germany}

\author{M. Xu}
\affiliation{Walther-Mei{\ss}ner-Institut, Bayerische Akademie der Wissenschaften, 85748 Garching, Germany}
\affiliation{Physik-Department, Technische Universit\"{a}t M\"{u}nchen, 85748 Garching, Germany}

\author{A. Parra-Rodriguez}
\affiliation{Department of Physical Chemistry, University of the Basque Country UPV/EHU, Apartado 644, E-48080 Bilbao, Spain}

\author{M. Sanz}
\affiliation{Department of Physical Chemistry, University of the Basque Country UPV/EHU, Apartado 644, E-48080 Bilbao, Spain}

\author{M. Fischer}
\affiliation{Walther-Mei{\ss}ner-Institut, Bayerische Akademie der Wissenschaften, 85748 Garching, Germany}
\affiliation{Physik-Department, Technische Universit\"{a}t M\"{u}nchen, 85748 Garching, Germany}
\affiliation{Nanosystems Initiative Munich (NIM), Schellingstra{\ss}e 4, 80799 M\"{u}nchen, Germany}

\author{E. Xie}
\affiliation{Walther-Mei{\ss}ner-Institut, Bayerische Akademie der Wissenschaften, 85748 Garching, Germany}
\affiliation{Physik-Department, Technische Universit\"{a}t M\"{u}nchen, 85748 Garching, Germany}
\affiliation{Nanosystems Initiative Munich (NIM), Schellingstra{\ss}e 4, 80799 M\"{u}nchen, Germany}

\author{K. Inomata}
\affiliation{RIKEN Center for Emergent Matter Science (CEMS), Wako, Saitama 351-0198, Japan}
\affiliation{National Institute of Advanced Industrial Science and Technology, 1-1-1 Umezono, Tsukuba, Ibaraki, 305-8563, Japan}

\author{Y. Nakamura}
\affiliation{RIKEN Center for Emergent Matter Science (CEMS), Wako, Saitama 351-0198, Japan}
\affiliation{Research Center for Advanced Science and Technology (RCAST), The University of Tokyo, Meguro-ku, Tokyo 153-8904, Japan}

\author{E. Solano}
\affiliation{Department of Physical Chemistry, University of the Basque Country UPV/EHU, Apartado 644, E-48080 Bilbao, Spain}
\affiliation{IKERBASQUE, Basque Foundation for Science, Maria Diaz de Haro 3, 48013, Bilbao, Spain}
\affiliation{Department of Physics, Shanghai University, 200444 Shanghai, China}

\author{A. Marx}
\affiliation{Walther-Mei{\ss}ner-Institut, Bayerische Akademie der Wissenschaften, 85748 Garching, Germany}

\author{F. Deppe}
\affiliation{Walther-Mei{\ss}ner-Institut, Bayerische Akademie der Wissenschaften, 85748 Garching, Germany}
\affiliation{Physik-Department, Technische Universit\"{a}t M\"{u}nchen, 85748 Garching, Germany}
\affiliation{Nanosystems Initiative Munich (NIM), Schellingstra{\ss}e 4, 80799 M\"{u}nchen, Germany}

\author{R. Gross}
\email[]{rudolf.gross@wmi.badw.de}
\affiliation{Walther-Mei{\ss}ner-Institut, Bayerische Akademie der Wissenschaften, 85748 Garching, Germany}
\affiliation{Physik-Department, Technische Universit\"{a}t M\"{u}nchen, 85748 Garching, Germany}
\affiliation{Nanosystems Initiative Munich (NIM), Schellingstra{\ss}e 4, 80799 M\"{u}nchen, Germany}

\begin{abstract}
Quantum communication protocols based on nonclassical correlations can be more efficient than known classical methods and offer intrinsic security over direct state transfer. In particular, remote state preparation aims at the creation of a desired and known quantum state at a remote location using classical communication and quantum entanglement. We present an experimental realization of deterministic continuous-variable remote state preparation in the microwave regime over a distance of \SI{35}{cm}. By employing propagating two-mode squeezed microwave states and feedforward, we achieve the remote preparation of squeezed states with up to 1.6\,dB of squeezing below the vacuum level. We quantify security in our implementation using the concept of the one-time pad. Our results represent a significant step towards microwave quantum networks between superconducting circuits.
\end{abstract}

\maketitle

In quantum technology, an efficient and secure exchange of quantum information between quantum nodes plays a crucial role\cite{cirac1997}. One of the first protocols realizing such a task was quantum teleportation, where an unknown quantum state is safely transferred from one party to another by using a shared entangled resource and classical feedforward\cite{Bennett1993,bouwmeester1997}. In a different scenario, where one party has full classical knowledge about a to-be-communicated quantum state, remote state preparation (RSP) can be used to remotely create this quantum state by employing similar tools as in quantum teleportation\cite{Bennett2001}. Compared to known classical methods, both protocols provide a quantum advantage as they require a smaller amount of classical information in the feedforward signal in order to communicate a desired quantum state\cite{kurucz2005}. However, in contrast to quantum teleportation, RSP allows for a nontrivial trade-off between the amount of classical communication and entanglement necessary for a successful protocol\cite{Bennett2001}. Furthermore, the use of an entangled resource allows RSP to operate perfectly secure\cite{kurucz2005}. Even though RSP is extensively investigated both theoretically and experimentally for discrete-variable systems\cite{Dakic2012,Peters2005,Ye2004}, deterministic implementations with continuous-variable systems are still lacking\cite{LeJeannic2018,Laurat2003}. At the same time, quantum communication based on continuous-variables is a field of intense research\cite{Weedbrook2012,Braunstein2005} investigating, e.g., quantum key distribution\cite{Jouguet2013}, quantum teleportation\cite{CVTele1998,Braunstein1998}, dense coding\cite{Mattle1996}, and free-space quantum communication\cite{Heim2014}.

Quantum communication in the microwave domain is motivated by the tremendous progress in the area of quantum information processing with superconducting circuits. In particular, the development of superconducting multi-qubit processors\cite{QSuprem2018,QED2016}, operated at gigahertz frequencies has been highly successful. We promote an approach of quantum communication directly in the microwave regime based on propagating squeezed states. Since these states have the same frequency and are generated by technology platforms already used for superconducting quantum computers, there is no mismatch between communication and data processing units. This approach is expected to be useful for short and medium distances, where superconducting waveguides can be used.

In this work, we realize deterministic continuous-variable RSP by creating Gaussian squeezed states over a distance of \SI{35}{cm}. We investigate the phase space of remotely preparable squeezed states and obtain good agreement with our model calculations based on the input-output formalism. Additionally, we find that our scheme corresponds to an extension of the one-time pad cryptographic protocol\cite{Vernam1926} into the quantum regime which allows for information-theoretic security. In contrast to already demonstrated quantum state transfer protocols between superconducting circuits\cite{Kurpiers2018,Axline2018}, our protocol does not directly transmit the target states to the receiving party. Moreover, it can be operated in the continuous regime and utilizes preshared entanglement to enable secure communication between parties. Since the generation and manipulation of Gaussian states is well understood\cite{Weedbrook2012}, they offer a viable option for building future intracity low-temperature quantum networks~\cite{Xiang2016}.

\begin{figure}
        \begin{center}
        \includegraphics[width=0.47\textwidth]{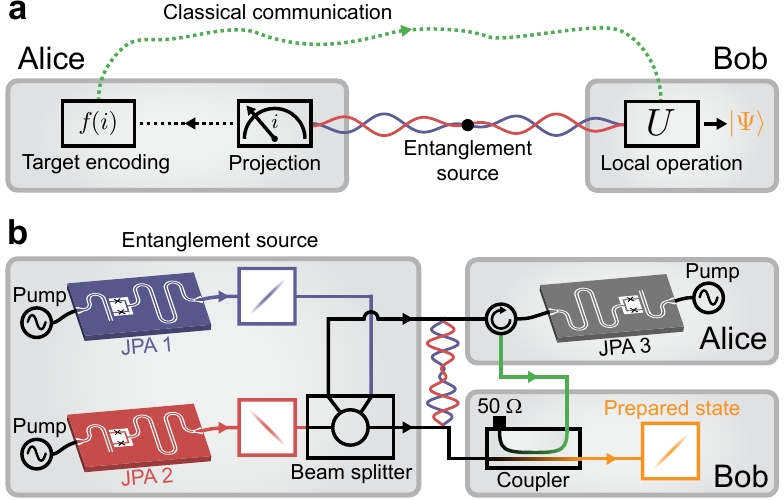}
        \end{center}
    \caption{\textbf{a} General RSP scheme: Alice remotely prepares a desired state at Bob's side using a quantum resource and classical communication (feedforward). \textbf{b} Experimentally implemented RSP scheme: a two-mode squeezed state (left) serves as quantum resource and the feedforward to Bob (right) is implemented using JPA\,3 and a directional coupler.}
    \label{fig1}
\end{figure}

\begin{figure*}[ht]
        \begin{center}
        \includegraphics{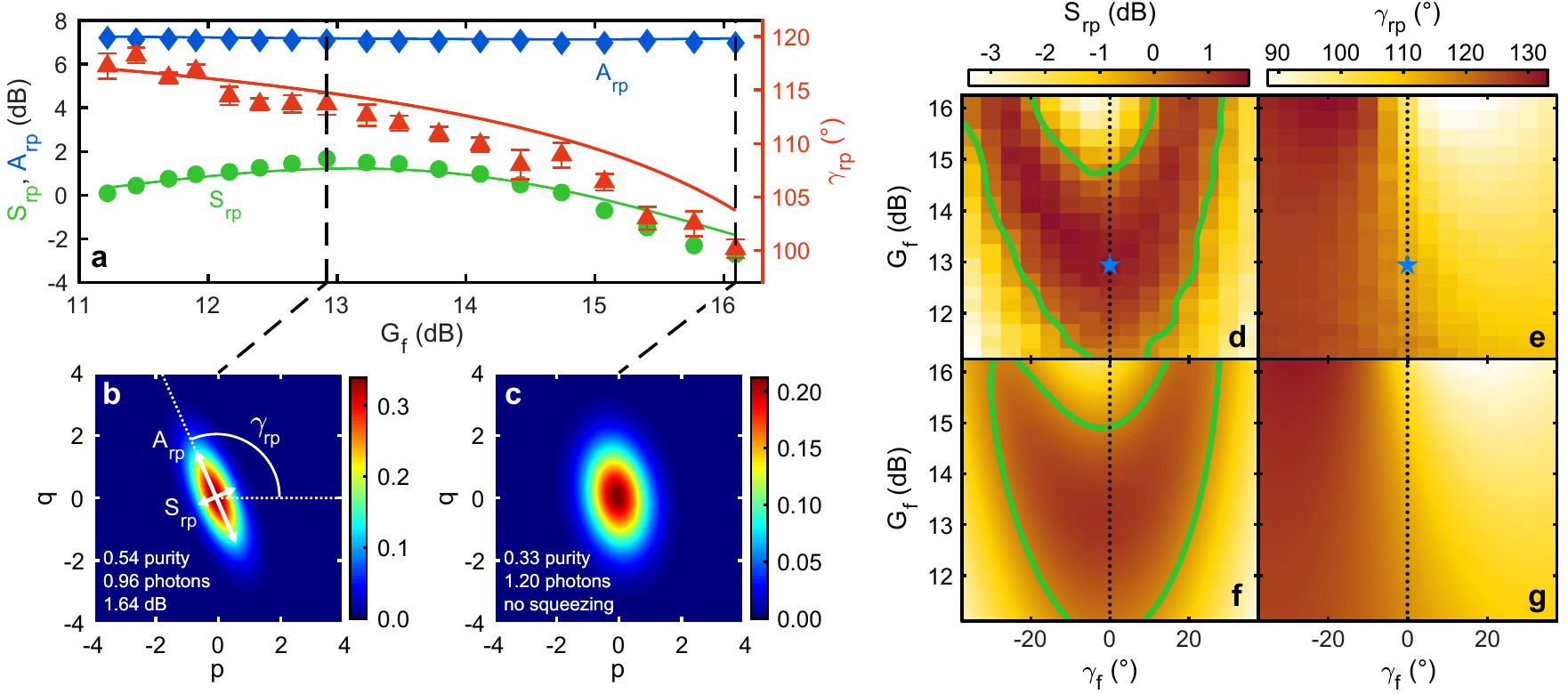}
        \end{center}
    \caption{\textbf{a} Squeezing level $S_\mathrm{rp}$ (circles), antisqueezing level $A_\mathrm{rp}$ (diamonds), and squeezing angle $\gamma_\mathrm{rp}$ (triangles) of remotely prepared states as a function of the JPA\,3 degenerate gain $G_\mathrm{f}$ at fixed angle $\gamma_\mathrm{f}\,{=}\,\SI{0}{\degree}$. The lines show a fit to the data. If not shown, the statistical errors are smaller than the symbol size. \textbf{b},\textbf{c} Reconstructed Wigner functions of the remotely prepared states for the optimal and one of the non-optimal JPA\,3 gains as marked by the dashed lines in panel \textbf{a}. \textbf{d},\textbf{e} $S_\mathrm{rp}$ and $\gamma_\mathrm{rp}$ of the remotely prepared states as a function of the feedforward parameters. Panels \textbf{f},\textbf{g} show a joint fit of the three quantities ($S_\mathrm{rp}$, $A_\mathrm{rp}$, $\gamma_\mathrm{rp}$) to the corresponding data in panels \textbf{d},\textbf{e}, respectively (see Supplementary Fig.\,3 for the results for $A_\mathrm{rp}$). The green lines mark the threshold $S_\mathrm{rp}\,{\ge}\,\SI{0}{dB}$ for squeezing below the vacuum limit. The optimal point is marked by the blue star. The data and fit in panel \textbf{a} are marked by dotted lines in panels \textbf{d}-\textbf{g}.}
    \label{fig2}
\end{figure*}

The general idea behind the RSP protocol and our experimental implementation using continuous-variable  microwave states are described in Fig.\,\ref{fig1}. We use flux-driven Josephson parametric amplifiers (JPAs) as the key elements for the generation and manipulation of squeezed microwave states\cite{Yamamoto2008,Pogorzalek2017,Goetz2016}. We operate all JPAs in the degenerate regime at the frequency $f_0 \,{=}\, \SI{5.435}{GHz}$ with a pump frequency $f_\mathrm{p} \,{=}\,2f_0$. The task of JPA\,1 and JPA\,2 is to generate propagating squeezed states which are incident to an entangling beam splitter.
The resulting symmetric two-mode squeezed (TMS)\cite{Fedorov2018,Flurin2012} states have a two-mode squeezing level\cite{Eichler2011b} of $S_\mathrm{TMS}\,{=}\, \SI{7.1}{dB}$ and an entanglement strength characterized by the negativity criterion\cite{Fedorov2018,Menzel2012} of $N\,{=}\,2.2$. This number quantifies the strength of nonlocal correlations present between field quadratures of signals propagating along different beam splitter output paths. Additionally, the symmetric TMS states have negligible local squeezing within each path. In other words, the microwave signals propagating on the two paths locally look like thermal states with, nevertheless, strong entanglement between them.

In the next step, we employ the symmetric propagating TMS states as a resource for remotely preparing the target squeezed microwave states. For this purpose, the TMS states are continuously distributed between two parties, Alice and Bob, who are separated by $\SI{35}{\cm}$ of superconducting cable. Alice generates a feedforward signal carrying the classical information about her choice on what quantum state is to be remotely prepared at Bob's side. Finally, Bob displaces his part of the resource state proportionally to the communicated signal by using a directional coupler with a fixed coupling of $\beta\,{\simeq}\,\SI{-15}{dB}$\cite{Fedorov2016,DiCandia2015}. We experimentally implement the feedforward by operating JPA\,3 as a phase-sensitive amplifier. Alice uses it to choose and strongly amplify a certain quadrature of the incoming TMS states. Note that, in contrast to the other JPAs, it does not matter whether the outgoing feedforward signal from JPA\,3 is squeezed or not~(Supplementary Note 5). The essential classical information, as required for a successful RSP, is encoded in the large instantaneous amplitude of the phase-sensitively amplified field quadrature. 

Figure\,\ref{fig2}a shows the experimental performance of the RSP scheme as a function of the JPA\,3 degenerate gain $G_\mathrm{f}$ for a fixed JPA\,3 amplification angle $\gamma_\mathrm{f} \,{=}\, \SI{0}{\degree}$. The latter is defined as the deviation from the angle of the optimal working point at which we achieve the highest purity in the remotely prepared states. We fully characterize these states in terms of their squeezing level $S_\mathrm{rp}$, antisqueezing level $A_\mathrm{rp}$, and squeezing angle $\gamma_\mathrm{rp}$ (see Methods). We clearly observe squeezing up to $S_\mathrm{rp}\,{=}\,\SI{1.6(1)}{dB}$ in the final states at the output of the displacer near the optimal JPA\,3 gain $G_\mathrm{f} \,{\simeq}\, \SI{13}{dB}$ (see Fig.\,\ref{fig2}b). $S_\mathrm{rp}$ decreases and the states even become non-squeezed upon deviation from the optimal JPA\,3 gain as shown in Fig.\,\ref{fig2}c. The remotely prepared states can be encoded not only by varying $G_\mathrm{f}$ but also by changing  $\gamma_\mathrm{f}$. The latter leads to a different quadrature in the resource TMS states being projected, and accordingly, to a different state being remotely prepared. The squeezing level and squeezing angle of the remotely prepared states obtained by sweeping both $G_\mathrm{f}$ and $\gamma_\mathrm{f}$ are shown in Fig.\,\ref{fig2}d,e. The results for the antisqueezing level $A_\mathrm{rp}$ can be found in Supplementary Fig.\,3.

Our experiment can be theoretically described by a model based on the input-output transformations for every component in the setup including transmission losses. Additionally, we assume imperfect JPAs adding a certain amount of noise with mean thermal photon numbers $n_i$  ($i\,{\in}\,\{1,2,\mathrm{f}\}$) to the JPA input signal. The RSP protocol is expected to work optimally for $G_\mathrm{f}\,{=}\,\tau/(1-\tau)$ and $\gamma_\mathrm{f}\,{=}\,\SI{0}{\degree}$, where $\tau\,{=}\,1-10^{\beta/10}$ is the transmissivity of the directional coupler. At this optimal point and under the condition $S_\mathrm{TMS}\,{\ge}\SI{3}{dB}$, we obtain for the squeezed variance of the remotely prepared state
\begin{equation}
\sigma_\mathrm{s}^2 = \frac{1}{4}\left[2(1+2n)e^{-2r}(1-\eta)\tau+2(\eta+n_\mathrm{f})\tau\right]\,,
\label{Eq-RSPoptimal}
\end{equation}
where we assume equal noise photon numbers $n_1\,{=}\,n_2\,{=}\,n$ and squeezing factors $r_1\,{=}\,r_2\,{=}\,r$ of JPA\,1 and JPA\,2 as well as equal losses $\eta$ after the beam splitter on both Alice's and Bob's side. Furthermore, $n_\mathrm{f}$ is the noise photon number of JPA\,3 (Supplementary Note 3).
Equation~(\ref{Eq-RSPoptimal}) indicates that the prepared squeezing level $S_\mathrm{rp}$ at the optimal point is at least \SI{3}{dB} below the squeezing level of the used resource.
In order to correctly model the experiment, we additionally include a finite crosstalk between JPA\,3 and the JPAs creating the TMS states as well as losses before the beam splitter. Figure\,\ref{fig2}f,g depicts a joint fit to the corresponding data. We observe a very good coincidence between the experimental results and our model for the following parameters: JPA\,1,2 squeezing levels $S\,{=}\,\SI{10.1}{dB}$, $n\,{=}\,0.04$, a total loss between JPA\,1,2 and the directional coupler input $\chi_1\,{=}\,\SI{1.0}{dB}$ and a total loss between JPA\,1,2 and JPA\,3 $\chi_2\,{=}\,\SI{1.1}{dB}$. All values nicely agree with those obtained from independent JPA characterization measurements and loss estimations.

The quantum advantage of the RSP protocol consists in a smaller amount of classical information sent through the feedforward channel in order to prepare a desired state as compared to a purely classical protocol\cite{kurucz2005,Killoran2010}. In the current experiment, this becomes evident by considering that only the amplified quadrature of the feedforward signal will affect the signal at Bob's side due to the low coupling $\beta\,{\ll}\,\SI{0}{dB}$ of the displacer. Consequently, we only send two real numbers while we are able to prepare different undisplaced mixed squeezed states which are fully described by three real numbers ($S_\mathrm{rp}$, $A_\mathrm{rp}$, $\gamma_\mathrm{rp}$).

\begin{figure}
	\begin{center}
	\includegraphics{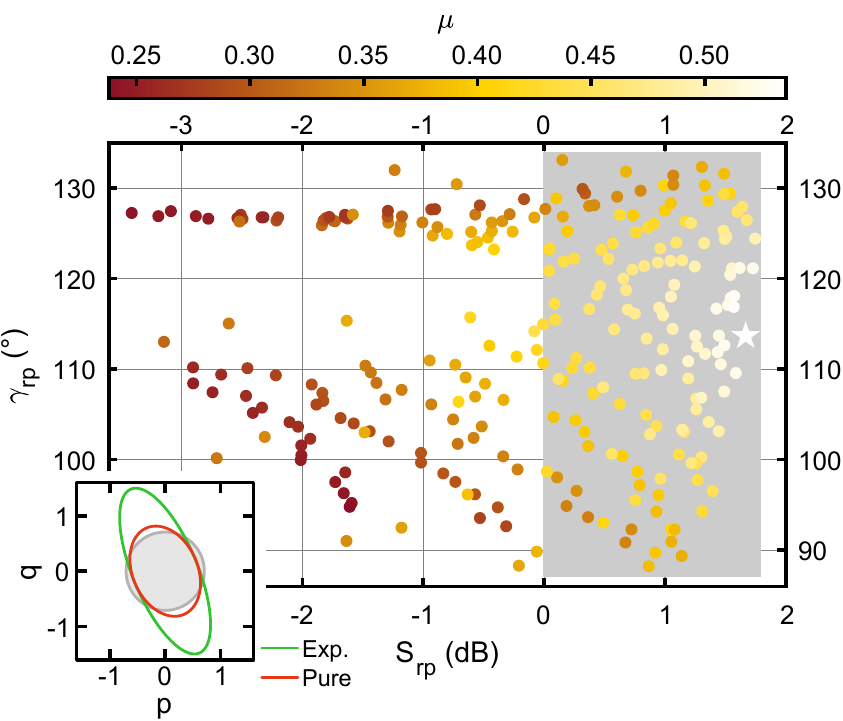}
	\end{center}
\caption{The phase space of the prepared states is spanned by $S_\mathrm{rp}$ and $\gamma_\mathrm{rp}$. The grey area marks squeezing below the vacuum limit. The color code indicates the purity $\mu$ of the remotely prepared states. The optimal point is marked by the white star. The inset shows the Wigner function $1/e$ contours of the experimental state at the optimal point with $\mu\,{=}\,0.54$ and a pure squeezed state with the same squeezing level. The vacuum $1/e$ contour is indicated in grey.}
\label{fig3}
\end{figure}

The manifold of undisplaced Gaussian states we can prepare is intuitively understood by plotting the results from Fig.\,\ref{fig2}d,e in the phase space of the prepared squeezing level and angle, as it is shown in Fig.\,\ref{fig3}. The purity $\mu\,{=}\,1/(4\sqrt{\mathrm{det}\sigma})$, where $\sigma$ is the covariance matrix of the remotely prepared state, incorporates the information about the antisqueezed quadrature and is a measure of how close the state is to a pure state. Compared to a perfectly prepared state with unit purity, we achieve the highest uncorrected purity $\mu\,{=}\,0.54\,{\pm}\,0.01$ at the optimal point which is sufficient for many applications of squeezed states such as entanglement generation\cite{Menzel2012}, sideband cooling of optomechanical systems\cite{Clark2017}, and quantum illumination\cite{LasHeras2017}. Using the model of the experiment, we identify that the observed purity is limited by the added noise of the JPAs and the losses in the setup. Upon reducing the JPA noise photon number by one order of magnitude as well as the total losses to $\chi_1\,{=}\,\chi_2\,{=}\,\SI{0.2}{dB}$, we expect an optimized purity $\mu_\mathrm{opt}\,{=}\,0.80$ for the prepared state at the optimal point. The reduction of losses can be achieved by using a superconducting hybrid ring, optimized cable connectors, and improved circulators. In this context, one should remember that our protocol allows for the preparation of continuous-variable squeezed states with a degree of squeezing $S_\mathrm{rp}$ that is fundamentally related to the initial two-mode squeezing of the resource state. In the current implementation, even for a fixed resource TMS state, $S_\mathrm{rp}$ and $\gamma_\mathrm{rp}$ can be changed at the expense of a reduced purity $\mu$. By adding a phase shifter\cite{Kokkoniemi2017} on her side, Alice could prepare squeezed states with arbitrary $\gamma_\mathrm{rp}$ while keeping $S_\mathrm{rp}$ and $\mu$ constant.

\begin{figure}
        \begin{center}
        \includegraphics{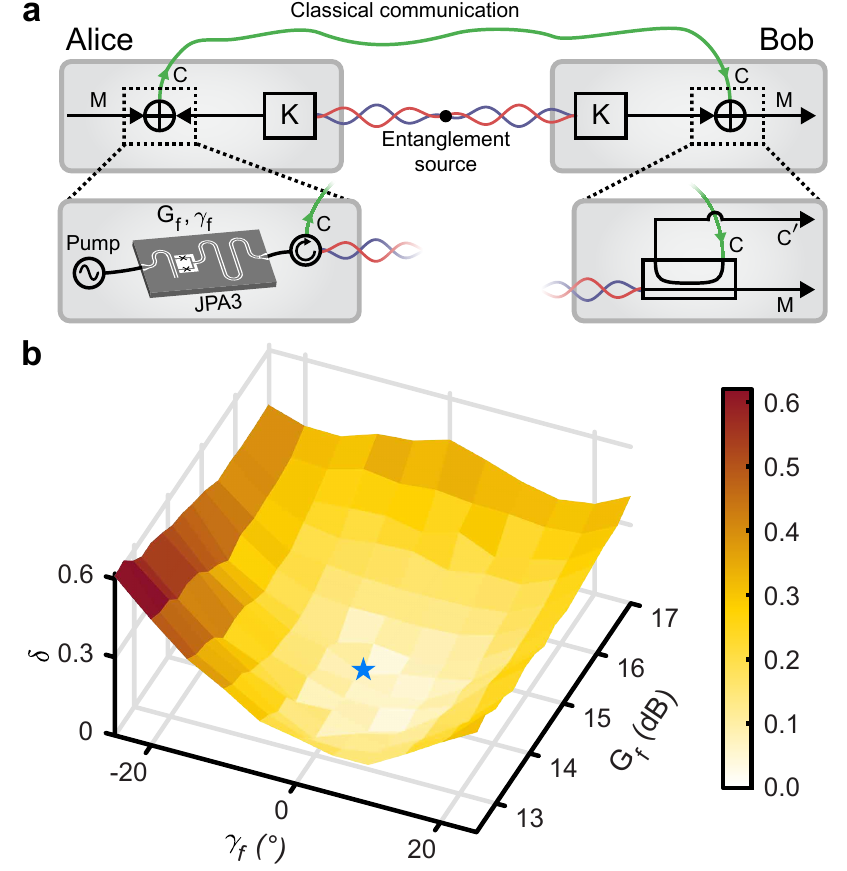}
        \end{center}
    \caption{\textbf{a} Scheme for the interpretation of the RSP setup in terms of the one-time pad. \textbf{b} Entropy difference $\delta\,{=}\,H(M)-H(M|C')$ as a function of the feedforward gain $G_\mathrm{f}$ and angle $\gamma_\mathrm{f}$. The optimal point is marked by the blue star. It is slightly shifted with respect to the previous measurements due to a modified cryogenic setup.}
    \label{fig4}
\end{figure}

Finally, we relate our experimental RSP scheme to the cryptographic protocol known as the one-time pad by extending the latter to the quantum regime~\cite{kurucz2005,Leung2002}. Here, Alice securely sends a quantum state $M$ to Bob over an insecure channel. We identify the transmitted message $M$ as the remotely prepared state on Bob's side and the openly communicated cipher $C$ as the feedforward signal (see Fig.~\ref{fig4}a). The entangled TMS states provide the random key $K$ in the form of quantum fluctuations to both parties. Note that $K$ is essential for the one-time pad since it is used by Alice and Bob to encode and decode $M$. For secure communication, $K$ needs to be a uniform random variable, such that an eavesdropper with knowledge about $C$ does not gain any information about $M$\cite{Shannon1949}. Formally, we can write
\begin{equation}\label{Eq:OTP}
H(M)-H(M|C)=0\, ,
\end{equation}
where $H(M)$ is the von Neumann entropy of the remotely prepared state and $H(M|C)\,{=}\,H(M,C)\,{-}\,H(C)$ is the conditional entropy of $M$ given the feedforward signal (Supplementary Note 6). We experimentally investigate the quantum one-time pad by measuring the prepared states as a function of the JPA\,3 parameters while additionally detecting the signal $C'$ from the second directional coupler output. We compute $\delta\,{=}\,H(M)-H(M|C')$ to verify equation~(\ref{Eq:OTP}) under the reasonable approximation $C'\,{\approx}\,C$ (due to $\tau\,{\simeq}\,1$) using state tomography. In Fig.~\ref{fig4}b, we observe a decrease in $\delta$ when moving towards the optimal point  where the smallest value $\delta\,{=}\,0.06\,{\pm}\,0.04$ is reached and the entropy of the prepared state is $H(M)\,{=}\,0.80\,{\pm}\,0.02$. The observation $\delta \ll H(M)$ indicates that the transfer of a quantum state from Alice to Bob is close to perfect security when approaching the optimal point.

To conclude, we have successfully implemented a deterministic RSP protocol over a distance of $\SI{35}{cm}$ in the microwave regime with continuous variables and explored the influence of different parameters on the remotely prepared states. We have remotely prepared squeezed states with a squeezing level of up to $\SI{1.6}{dB}$ below the vacuum limit. In our specific RSP implementation, Alice can control the squeezing level and, to some extent, the squeezing angle of the remotely prepared state at the expense of a reduced purity. Additionally, we demonstrate that the protocol can be interpreted as a secure one-time pad near the optimal point. The operational range of both the RSP and quantum one-time pad protocols can be extended to any angle $\gamma_\mathrm{rp}$ by an additional phase shifter on Alice's side. The demonstrated protocol opens a way to a multitude of intriguing experiments with quantum microwaves such as probing the Holevo bound limits\cite{Zwolak2013}, studying the role of quantum discord in quantum communication protocols\cite{Gu2012}, exploring hybrid continuous-discrete schemes of quantum information processing\cite{Andersen2015}, and implementing quantum illumination protocols\cite{LasHeras2017}. Our experiment proves that prototypical local quantum networks using continuous-variable quantum microwaves are within experimental reach.

\medskip\noindent
\textbf{\large Methods}

\noindent
JPA\,1 and JPA\,2 perform a squeezing operation $\hat{S}(\xi) |0\rangle$, where $\hat{S}(\xi) \,{=}\, \exp(\frac{1}{2}\xi^* \hat{a}^2 \,{-}\, \frac{1}{2}\xi (\hat{a}^\dagger)^2)$ is the squeezing operator, $\hat{a}^{\dagger}\,{=}\,\hat{q}-i\hat{p}$ and $\hat{a}\,{=}\,\hat{q}+i\hat{p}$ are the creation and annihilation operators with $\left[\hat{a},\hat{a}^\dagger\right]\,{=}\,1$ of the $f_0$ mode with quadratures $\hat{q}$ and $\hat{p}$, and $\xi \,{=}\, r e^{i\phi}$ is the complex squeezing amplitude.  Here, the phase $\phi \,{=-}2\gamma$ determines the squeezing angle $\gamma$ between the antisqueezed quadrature and the $p$-axis in the phase space, while the squeezing factor $r$ parameterizes the amount of squeezing. We define the degree of squeezing in decibels as $S \,{=}\,{-}10\,\log_{10} [\sigma_\mathrm{s}^2 / 0.25]$, where $\sigma_\mathrm{s}^2$ is the variance of the squeezed quadrature and the vacuum variance is $0.25$. Positive values of $S$ indicate squeezing below the vacuum level. The antisqueezing level is defined as $A\,{=}\,10\,\log_{10} [\sigma_\mathrm{a}^2 / 0.25]$, where $\sigma_\mathrm{a}^2$ is the variance of the antisqueezed quadrature. We generate symmetric two-mode squeezed states at the output of the hybrid ring by pumping JPA\,1 and JPA\,2 with strong quasi-continuous microwave drives so that they produce squeezed states with the same squeezing level and orthogonal squeezing angles $\gamma_2\,{=}\,\gamma_1\,{+}\,\pi/2$. These angles are stabilized by controlling the respective pump phases employing a phase-locked loop\cite{Fedorov2016,Fedorov2018}. The stability of these TMS states in terms of two-mode squeezing and symmetry is of paramount importance in our experiments. Only by utilizing the nonclassical correlations between Alice and Bob, it is possible to demonstrate the successful RSP protocol.
In order to reconstruct the quantum states in the experiment, we employ a well-tested reference state tomography based on statistical moments of the detected field quadratures\cite{Menzel2012,Eichler2011b}.

\medskip\noindent
\textbf{\large Acknowledgments}

\noindent
We acknowledge support by the German Research Foundation through FE 1564/1-1 and the Munich Center for Quantum Science and Technology (MCQST), Elite Network of Bavaria through the program ExQM, EU Flagship project QMiCS and OpenSuperQ, Spanish MINECO/FEDER FIS2015-69983-P, Basque Government Grant No. IT986-16 and PhD Grant No. PRE-2016-1-0284, projects JST ERATO (Grant No. JPMJER1601) and JSPS KAKENHI (Grant No. 26220601 and Grant No. 15K17731). This material is also based upon work supported by the U.S. Department of Energy, Office of Science, Office of Advance Scientific Computing Research (ASCR), under field work proposal number ERKJ335. We would like to thank K. Kusuyama for assistance with part of the JPA fabrication and S. B. Ghaffari for help in early stages of the experiment.

\bibliographystyle{naturemag}
\bibliography{Bibliography_RSP}
\end{document}


\title[]{Supplementary Information: \\ Secure quantum remote state preparation of squeezed microwave states}

\author{S.~Pogorzalek}
\email[]{stefan.pogorzalek@wmi.badw.de}
\affiliation{Walther-Mei{\ss}ner-Institut, Bayerische Akademie der Wissenschaften, 85748 Garching, Germany}
\affiliation{Physik-Department, Technische Universit\"{a}t M\"{u}nchen, 85748 Garching, Germany}

\author{K. G. Fedorov}
\email[]{kirill.fedorov@wmi.badw.de}
\affiliation{Walther-Mei{\ss}ner-Institut, Bayerische Akademie der Wissenschaften, 85748 Garching, Germany}
\affiliation{Physik-Department, Technische Universit\"{a}t M\"{u}nchen, 85748 Garching, Germany}

\author{M. Xu}
\affiliation{Walther-Mei{\ss}ner-Institut, Bayerische Akademie der Wissenschaften, 85748 Garching, Germany}
\affiliation{Physik-Department, Technische Universit\"{a}t M\"{u}nchen, 85748 Garching, Germany}

\author{A. Parra-Rodriguez}
\affiliation{Department of Physical Chemistry, University of the Basque Country UPV/EHU, Apartado 644, E-48080 Bilbao, Spain}

\author{M. Sanz}
\affiliation{Department of Physical Chemistry, University of the Basque Country UPV/EHU, Apartado 644, E-48080 Bilbao, Spain}

\author{M. Fischer}
\affiliation{Walther-Mei{\ss}ner-Institut, Bayerische Akademie der Wissenschaften, 85748 Garching, Germany}
\affiliation{Physik-Department, Technische Universit\"{a}t M\"{u}nchen, 85748 Garching, Germany}
\affiliation{Nanosystems Initiative Munich (NIM), Schellingstra{\ss}e 4, 80799 M\"{u}nchen, Germany}

\author{E. Xie}
\affiliation{Walther-Mei{\ss}ner-Institut, Bayerische Akademie der Wissenschaften, 85748 Garching, Germany}
\affiliation{Physik-Department, Technische Universit\"{a}t M\"{u}nchen, 85748 Garching, Germany}
\affiliation{Nanosystems Initiative Munich (NIM), Schellingstra{\ss}e 4, 80799 M\"{u}nchen, Germany}

\author{K. Inomata}
\affiliation{RIKEN Center for Emergent Matter Science (CEMS), Wako, Saitama 351-0198, Japan}
\affiliation{National Institute of Advanced Industrial Science and Technology, 1-1-1 Umezono, Tsukuba, Ibaraki, 305-8563, Japan}

\author{Y. Nakamura}
\affiliation{RIKEN Center for Emergent Matter Science (CEMS), Wako, Saitama 351-0198, Japan}
\affiliation{Research Center for Advanced Science and Technology (RCAST), The University of Tokyo, Meguro-ku, Tokyo 153-8904, Japan}

\author{E. Solano}
\affiliation{Department of Physical Chemistry, University of the Basque Country UPV/EHU, Apartado 644, E-48080 Bilbao, Spain}
\affiliation{IKERBASQUE, Basque Foundation for Science, Maria Diaz de Haro 3, 48013, Bilbao, Spain}
\affiliation{Department of Physics, Shanghai University, 200444 Shanghai, China}

\author{A. Marx}
\affiliation{Walther-Mei{\ss}ner-Institut, Bayerische Akademie der Wissenschaften, 85748 Garching, Germany}

\author{F. Deppe}
\affiliation{Walther-Mei{\ss}ner-Institut, Bayerische Akademie der Wissenschaften, 85748 Garching, Germany}
\affiliation{Physik-Department, Technische Universit\"{a}t M\"{u}nchen, 85748 Garching, Germany}
\affiliation{Nanosystems Initiative Munich (NIM), Schellingstra{\ss}e 4, 80799 M\"{u}nchen, Germany}

\author{R. Gross}
\email[]{rudolf.gross@wmi.badw.de}
\affiliation{Walther-Mei{\ss}ner-Institut, Bayerische Akademie der Wissenschaften, 85748 Garching, Germany}
\affiliation{Physik-Department, Technische Universit\"{a}t M\"{u}nchen, 85748 Garching, Germany}
\affiliation{Nanosystems Initiative Munich (NIM), Schellingstra{\ss}e 4, 80799 M\"{u}nchen, Germany}

\date{\today}

\maketitle

\widetext
\setcounter{equation}{0}
\setcounter{figure}{0}
\setcounter{table}{0}
\makeatletter
\renewcommand{\theequation}{S\arabic{equation}}
\renewcommand{\thefigure}{S\arabic{figure}}
\renewcommand{\thetable}{\arabic{table}}

\setcounter{secnumdepth}{0}

\section{Josephson Parametric Amplifiers}

\noindent
The Josephson parametric amplifiers (JPA) used in this work consist of a quarter-wavelength superconducting microwave resonator in a coplanar waveguide geometry which is short-circuited to the ground plane via a direct current superconducting quantum interference device (dc-SQUID)\cite{Yamamoto2008,Pogorzalek2017}. The JPAs were designed and fabricated at NEC Smart Energy Research Laboratories, Japan and RIKEN, Japan. The resonator and pump line are patterned into a \SI{50}{nm} thick Nb film which has been deposited by magnetron sputtering onto $\SI{300}{\mu m}$ thick silicon substrates covered by a thermal oxide. The dc-SQUID is fabricated using an aluminum shadow evaporation technique. The flux-tunable resonant frequency $f_0$ of the JPA can be tuned by an external magnetic flux applied to the dc-SQUID loop via an external coil or via an on-chip antenna acting as the pump line. In order to squeeze incoming vacuum fluctuations or perform phase-sensitive amplification of the mode $f_0$, we apply a strong coherent pump tone at frequency $f_\mathrm{p}\,{=}\,2f_0$ to the pump line. The squeezing strength (degenerate gain) and squeezing angle (amplified quadrature) are controlled by the pump amplitude and pump phase, respectively, when the JPA is operated as squeezer (degenerate amplifier). For each JPA, a commercial cryogenic circulator is used to separate the incoming from the outgoing signal (see Fig.~\ref{fig:ExpSetup}). 
In order to pre-characterize the JPAs and find a suitable working frequency $f_0$ for all JPAs, we perform spectroscopic measurements\cite{Pogorzalek2017}. The extracted parameters are summarized in Tab.~\ref{tab:JPA_params}.

\begin{table*}[h]
\caption{JPA Parameters extracted by fitting the dependence of the resonant frequency $f_0$ of the JPAs on the applied magnetic flux\cite{Pogorzalek2017}. Here, $I_c$ and $E_\mathrm{J}\,{=}\,I_c\Phi_0/2\pi$ are the critical current and coupling energy of a single Josephson junction, respectively, $L_\mathrm{loop}$ and $\beta_\mathrm{L}\,{=}\,2L_\mathrm{loop}I_c/\Phi_0$ are the loop inductance and screening parameter of the dc-SQUID, respectively, and $f_r$ is the resonant frequency of the bare resonator. The Josephson junctions of the dc-SQUID are assumed to be identical. The external quality factors $Q_\mathrm{ext}$ and internal quality factors $Q_\mathrm{int}$ are obtained from independent fits of the JPA spectral linewidths\cite{Pogorzalek2017}. The parameters of JPA\,3 are similar to the ones of JPA\,1 and JPA\,2.}
\begin{ruledtabular}
\begin{tabular}{l*{8}{c}r}
Sample            & $I_\mathrm{c}$ ($\mu$A) & $\beta_\mathrm{L}$& $L_\mathrm{loop}$ (pH) & $f_\mathrm{r}/2\pi$ (GHz) & $E_\mathrm{J}/h$ ($\mathrm{THz}$) & $Q_\mathrm{ext}$ & $Q_\mathrm{int}$ \\
\hline
JPA\,1	  & 2.45 & 0.09 & 35.8 & 5.808 & 1.22 & 300--360 & $>$30000 \\
JPA\,2	  & 2.41 & 0.10 & 40.7 & 5.838 & 1.20 & 240--260 & $>$30000 \\
\end{tabular}
\end{ruledtabular}
\label{tab:JPA_params}
\end{table*}

\section{Experimental setup}

\begin{figure}[ht]
\includegraphics[width=0.88\textwidth]{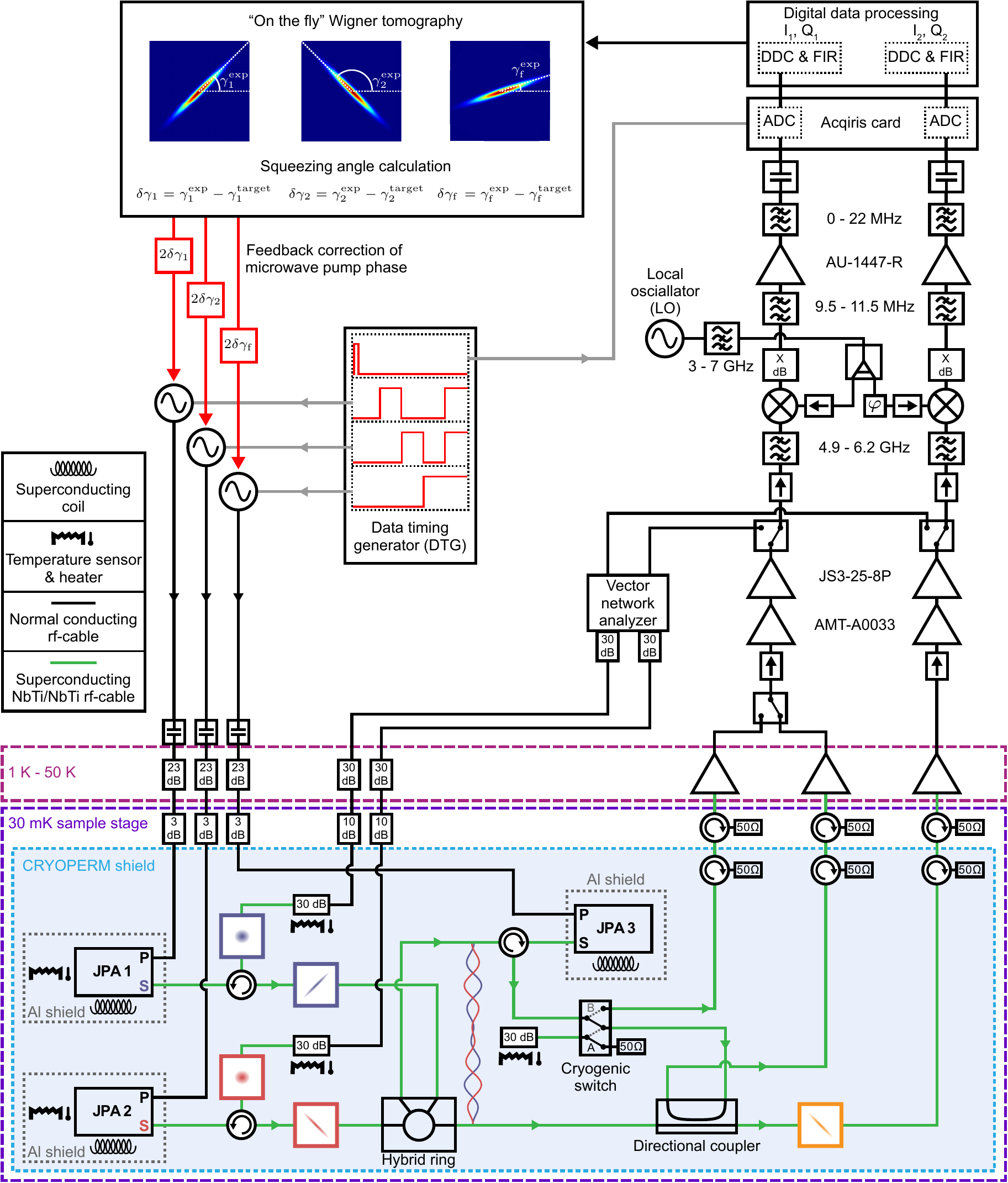}
\caption{Experimental scheme for the measurements. The JS3-25-8P rf-amplifiers are removed for the measurements concerning the quantum one-time pad. The RSP and quantum one-time pad measurements are performed with the cryogenic switch in position A. The intertwined lines between the outputs of the hybrid ring symbolize the entanglement. JPA\,3 and the directional coupler are separated by \SI{35}{cm} of superconducting cable.}
\label{fig:ExpSetup}
\end{figure}

\noindent
The experimental room temperature and cryogenic setup is shown in Fig.~\ref{fig:ExpSetup}. The digitizer card and the microwave pump sources for each JPA are pulsed with a data timing generator (DTG). JPA\,1 and JPA\,2 are both temperature stabilized at \SI{50}{mK} in order to ensure a stable JPA operation and produce squeezed states with orthogonal squeezing angles. The two squeezed states are superimposed by a cryogenic hybrid ring (50:50 beam splitter) in order to produce path-entangled two-mode squeezed (TMS) states at the outputs of the hybrid ring. By operating JPA\,1 and JPA\,2 at the same squeezing level, we are able to produce symmetric TMS states with local statistics of a thermal state. One output path of the beam splitter is connected to JPA\,3 which is operated as a phase-sensitive amplifier. The JPA\,3 output signal is then either detected or sent to a directional coupler which couples to the other hybrid ring output. The first amplification stage of a high-electron-mobility transistor (HEMT) is followed by additional rf-amplifiers which are temperature stabilized with a Peltier cooler. We use a vector network analyzer for the characterization of the JPAs and a heterodyne detection setup for the tomographic measurements.

The heterodyne detection setup and data processing is similar to those described in Refs.~\onlinecite{Fedorov2018,Menzel2010} where the signal is roughly filtered around the working frequency and down-converted to \SI{11}{MHz} by image rejection mixers. The signal is then digitized with analog-to-digital (ADC) converters on an Acqiris DC440 card. After sending the digitized data to a computer, digital data processing is performed where digital down-conversion (DDC) and finite-impulse response (FIR) filtering with a full bandwidth of \SI{430}{kHz} is applied. Finally, all correlation quadrature moments $\langle I_1^n I_2^m Q_1^k Q_2^l\rangle$ with $n\,{+}\,m\,{+}\,k\,{+}\,l\,{\leq}\,4$ for $n,m,k,l\in \mathbb{N}$ are calculated and averaged. The data within a single averaging cycle consists of $~4\times10^8$ averaged sample points per part of the pulse and is used to perform a reference state reconstruction for each pulse in order to obtain the signal moments $\langle(\hat{a}^\dagger)^n\hat{a}^m\rangle$ with $n\,{+}\,m\,{\leq}\,4$. During each measurement cycle, the moments of JPAs\,1-3 are used to calculate the squeezing angles $\gamma_i^\mathrm{exp}$ for each JPA ``on the fly" in order to obtain the angle correction $\delta\gamma_i\,{=}\,\gamma_i^\mathrm{ext}-\gamma_i^\mathrm{target}$ which are used to adjust the phase of the microwave pump tone by $2\delta\gamma_i$. Finally, the described averaging cycle is repeated 10 times. The vector network analyzer, DTG, Acqiris card and local oscillator are synchronized to a \SI{10}{MHz} rubidium frequency standard. The pump microwave sources are daisy chained to the local oscillator with a \SI{1}{GHz} reference signal.

The experimental states are reconstructed under the assumption that the states are Gaussian, and thus fully described by their signal moments up to the second order. In order to check for the Gaussianity of the states, we verify that the cumulants of third and fourth order are vanishingly small, as expected for Gaussian states\cite{Menzel2012}. The cumulants $\langle\langle(\hat{a}^\dagger)^n\hat{a}^m\rangle\rangle$ are calculated from the signal moments $\langle(\hat{a}^\dagger)^n\hat{a}^m\rangle$ according to
\begin{equation}
\langle\langle(\hat{a}^\dagger)^n\hat{a}^m\rangle\rangle = \partial^n_x \partial^m_y \mathrm{ln} \sum_{\alpha,\beta} \left. \frac{\langle (\hat{a}^\dagger)^\alpha\hat{a}^\beta \rangle x^\alpha y^\beta}{\alpha !\beta !}\right\rvert_{x=y=0}\,,
\end{equation}
where $\partial^n_x$ is the n-th partial derivative with respect to $x$ and ln is the natural logarithm\cite{Eichler2014,Xiang2018}.
 
\section{Theoretical modeling and fitting procedure}\label{sec:theory}

\begin{figure}[h]
\includegraphics{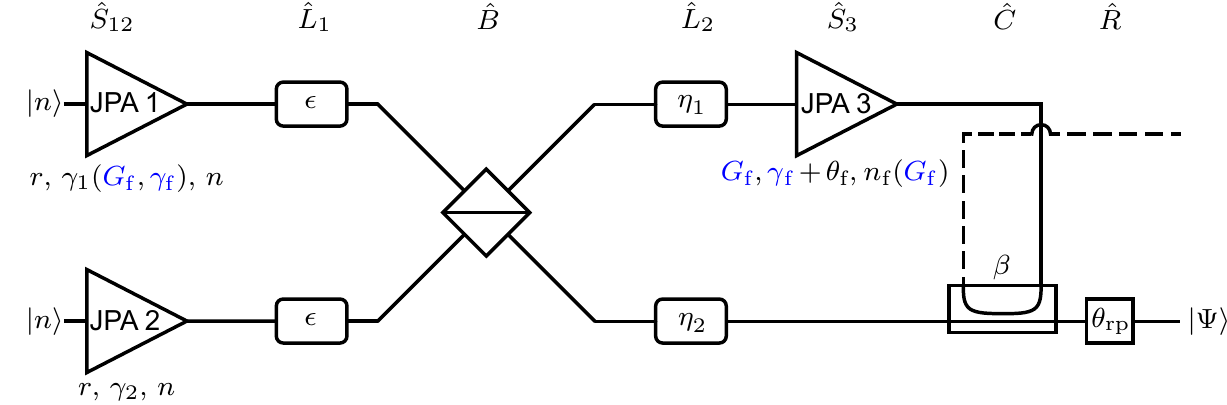}
\caption{Scheme for theoretical description of the RSP setup with used parameters. The experimentally varied feedforward gain $G_\mathrm{f}$ and angle $\gamma_\mathrm{f}$ are marked in blue. Due to a carefully designed symmetric implementation in the experiment, the losses $\epsilon_1\,{=}\,\epsilon_2\,{=}\,\epsilon$ in both paths before the beam splitter are assumed to be equal and include the insertion loss of the beam splitter. The angles $\gamma_1$, $\gamma_2$ and $\gamma_\mathrm{f}$ are used in units of radians for all equations.}
\label{fig:RSP_Scheme}
\end{figure}

\noindent
In order to theoretically describe the remote state preparation (RSP) setup, we use an input-output model for each component as shown in Fig.\,\ref{fig:RSP_Scheme}. JPA\,1 and JPA\,2 are modeled as squeezers with the same squeezing parameter $r_1\,{=}\,r_2\,{=}\,r$ but different squeezing angles $\gamma_{1}$ and $\gamma_{2}$ in order to produce symmetric TMS states after the beam splitter. In the experiment, the pump of JPA\,3 leaks through to JPA\,1 and JPA\,2 which results in a finite crosstalk between JPA\,3 and the other two JPAs. Since experimentally the crosstalk to JPA\,1 dominates, we approximate the effect of the crosstalk by a linear dependence of $\gamma_{1}$ on the gain $G_\mathrm{f}$ and angle $\gamma_\mathrm{f}$ of JPA\,3
\begin{equation}
 \gamma_{1}=\gamma_1^{(0)} + \kappa G_\mathrm{f} + \lambda \gamma_\mathrm{f} \,,
 \label{eqn:gamma1}
\end{equation}
where $\gamma_1^{(0)}$ is the unperturbed squeezing angle of JPA\,1. This approximation is consistent with independent measurements of the crosstalk. The squeezing operator $\hat{S}_{12}$ for JPA\,1 and JPA\,2 acting on the annihilation operators $\hat{a}_i$ of path 1 (Alice) and path 2 (Bob) is given by\cite{Scully1997}
\begin{equation}
\hat{S}_{12}^\dagger \begin{pmatrix}  \hat{a}_1   \\ \hat{a}_2  \end{pmatrix} \hat{S}_{12} =  \begin{pmatrix} \hat{a}_1\mathrm{cosh}\,r - \hat{a}_1^\dagger e^{-2i\gamma_1} \mathrm{sinh}\,r  \\ \hat{a}_2\mathrm{cosh}\,r - \hat{a}_2^\dagger e^{-2i\gamma_2} \mathrm{sinh}\,r  \end{pmatrix} \, .
\label{eqn:S12}
\end{equation}
The added noise of JPA\,1 and JPA\,2 is taken into account by an effective thermal state with a noise photon number $n_1\,{=}\,n_2\,{=}\,n$ incident to the JPAs. 
In order to describe the action of JPA\,3, we assume that classical noise is added to the JPA input signal followed by ideal phase-sensitive amplification
\begin{equation}
\hat{S}_{3}^\dagger \begin{pmatrix}  \hat{a}_1   \\ \hat{a}_2  \end{pmatrix} \hat{S}_{3} =  \begin{pmatrix} \left(\hat{a}_1+\zeta\right)\mathrm{cosh}\,r_\mathrm{f} - \left(\hat{a}_1^\dagger+\zeta^*\right) e^{-2i(\gamma_\mathrm{f}+\theta_\mathrm{f})} \mathrm{sinh}\,r_\mathrm{f}  \\ \hat{a}_2 \end{pmatrix} \, ,
\end{equation}
where $G_\mathrm{f}$ is related to $r_\mathrm{f}$ as $G_\mathrm{f}\,{=}\,e^{2r_\mathrm{f}}$ and $\theta_\mathrm{f}$ is the theoretically optimal JPA\,3 amplification angle. The classical noise is described by the complex Gaussian random variable $\zeta$ with $\langle\zeta\rangle\,{=}\,0$, $\langle\zeta\zeta^*\rangle\,{=}\,n_\mathrm{f}$ and $\langle \mathrm{Re}(\zeta)^2\rangle\,{=}\,\langle \mathrm{Im}(\zeta)^2\rangle\,{=}\,n_\mathrm{f}/2$, where $n_\mathrm{f}$ is the effective thermal noise photon number. In general, the JPA noise is gain dependent which we take into account by a linear dependence on $G_\mathrm{f}$ for JPA\,3. For that, we use $n_\mathrm{f}\,{=}\,n'_\mathrm{f}G_\mathrm{f}$, where $n'_\mathrm{f}$ is a proportionality constant.

Losses $\epsilon$ and $\eta_i$ of the microwave components are modeled with a beam splitter\cite{DiCandia2015}
\begin{equation}
\hat{L}_1^\dagger \begin{pmatrix}  \hat{a}_1   \\ \hat{a}_2  \end{pmatrix} \hat{L}_1 =  \begin{pmatrix} \sqrt{1-\epsilon}\hat{a}_1 + \sqrt{\epsilon}\hat{v}_1  \\ \sqrt{1-\epsilon}\hat{a}_2 + \sqrt{\epsilon}\hat{v}_2  \end{pmatrix} \, ,
\end{equation}
\begin{equation}
\hat{L}_2^\dagger \begin{pmatrix}  \hat{a}_1   \\ \hat{a}_2  \end{pmatrix} \hat{L}_2 =  \begin{pmatrix} \sqrt{1-\eta_1}\hat{a}_1 + \sqrt{\eta_1}\hat{v}_1  \\ \sqrt{1-\eta_2}\hat{a}_2 + \sqrt{\eta_2}\hat{v}_2  \end{pmatrix} \, ,
\end{equation}
where $\hat{v}_i$ is the operator describing the environment for path $i$. The environment can be safely approximated to be in the vacuum state due to the low temperature of the lossy components in the experiment.

The hybrid ring is described by a 50:50 beam splitter\cite{Braunstein2005}
\begin{equation}
\hat{B}^\dagger \begin{pmatrix}  \hat{a}_1   \\ \hat{a}_2  \end{pmatrix} \hat{B} =  \frac{1}{\sqrt{2}}\begin{pmatrix} \hat{a}_1 + \hat{a}_2 \\ -\hat{a}_1 + \hat{a}_2  \end{pmatrix} \, .
\end{equation}

The displacement on Bob's side is implemented with a directional coupler and is described as an asymmetric beam splitter\cite{Paris1996}
\begin{equation}
\hat{C}^\dagger \begin{pmatrix}  \hat{a}_1   \\ \hat{a}_2  \end{pmatrix} \hat{C} =  \begin{pmatrix} \sqrt{\tau}\hat{a}_1 + \sqrt{1-\tau}\hat{a}_2 \\ -\sqrt{1-\tau}\hat{a}_1 + \sqrt{\tau}\hat{a}_2  \end{pmatrix} \, ,
\end{equation}
where $\tau=1-10^{\beta/10}$ is the transmissivity and $\beta$ is the coupling in decibel.

\begin{figure}[t]
\includegraphics[width=0.7\textwidth]{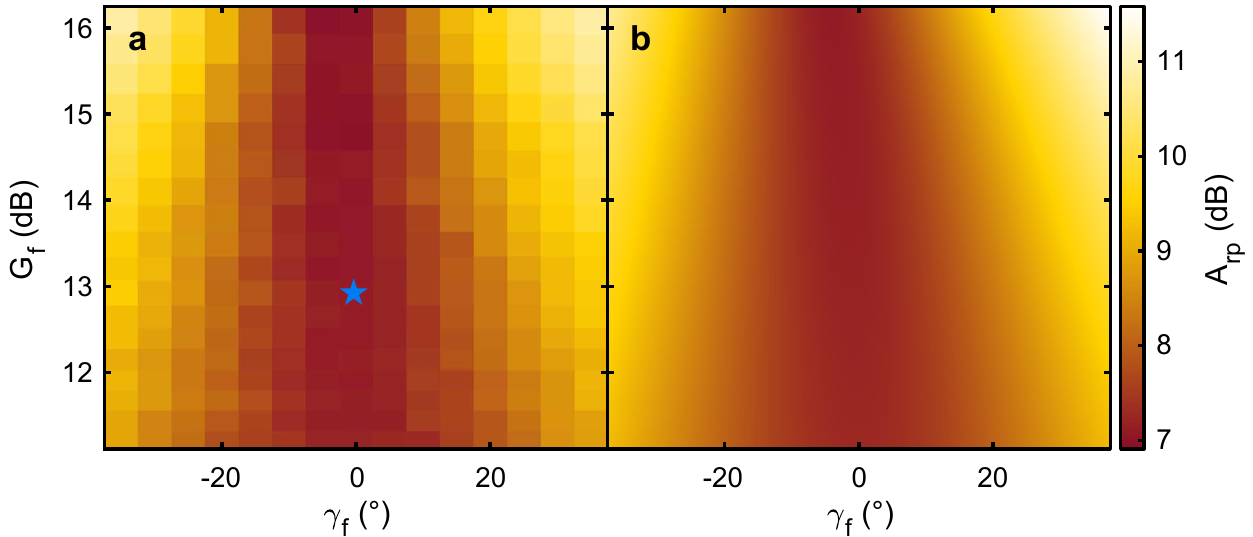}
\caption{\textbf{a} and \textbf{b}, Antisqueezed variance $A_\mathrm{rp}$ of the remotely prepared states as a function of the feedforward parameters for experiment and fit, respectively. The optimal point is marked by the blue star.}
\label{fig:AntiSq}
\end{figure}

In order to describe the realistic setup, we need to take the electrical length of the different components into account. The total electrical lengths as well as different path lengths after the beam splitter are compensated with a rotation~$\hat{R}$ by the angle $\theta_\mathrm{rp}$ of the final remotely prepared state on Bob's side
\begin{equation}
\hat{R}^\dagger \begin{pmatrix}  \hat{a}_1   \\ \hat{a}_2  \end{pmatrix} \hat{R} =  \begin{pmatrix} \hat{a}_1 \\ \hat{a}_2 e^{-i\theta_\mathrm{rp}}  \end{pmatrix} \, .
\label{eqn:R}
\end{equation}
With the operator definitions in equations~(\ref{eqn:S12})--(\ref{eqn:R}), we can write the overall RSP protocol as
\begin{equation}
|\Psi\rangle=\hat{R} \, \hat{C} \, \hat{S}_3 \, \hat{L}_2 \, \hat{B} \, \hat{L}_1 \, \hat{S}_{12}|n,n\rangle\, ,
\end{equation}
where $n$ is the noise photon number of JPA\,1 and JPA\,2, and $|\Psi\rangle$ is the final state on both paths. The moment matrices for both paths of the final state are calculated as
\begin{equation}\label{eq:RSPstate}
\begin{pmatrix} \langle(\hat{b}^\dagger)^n \hat{b}^m \rangle_1 \\ \langle(\hat{b}^\dagger)^n \hat{b}^m \rangle_2 \end{pmatrix}=\langle\Psi|\begin{pmatrix}(\hat{a}^\dagger)_1^n \hat{a}_1^m \\ (\hat{a}^\dagger)_2^n \hat{a}_2^m \end{pmatrix} |\Psi\rangle\, ,
\end{equation}
where $\langle(\hat{b}^\dagger)^n \hat{b}^m \rangle_1$ are the moments of the second directional coupler output signal and $\langle(\hat{b}^\dagger)^n \hat{b}^m \rangle_2$ are the moments of the remotely prepared  state. With the definition of the quadratures $\hat{q}=(\hat{b}+\hat{b}^\dagger)/2$ and $\hat{p}=(\hat{b}-\hat{b}^\dagger)/2i$, the moments $\langle(\hat{b}^\dagger)^n \hat{b}^m \rangle_2$  are used to calculate the squeezing angle~$\gamma_\mathrm{rp}$, squeezed variance~$\sigma_\mathrm{s}^2$ and antisqueezed variance~$\sigma_\mathrm{a}^2$ of the remotely prepared state as
\begin{align}
\gamma_\mathrm{rp} &= -\frac{1}{2}\mathrm{arg}\left(-\langle\hat{b}^2 \rangle_2\right)\, , \label{eq:RSPgamma}\\
\sigma_\mathrm{s}^2 &= \frac{1}{4}\left( \langle\hat{b}^2 \rangle_2 e^{2i\gamma_\mathrm{rp}}+\langle(\hat{b}^\dagger)^2 \rangle_2 e^{-2i\gamma_\mathrm{rp}}+2\langle\hat{b}^\dagger\hat{b}\rangle_2+1 \right)\, , \label{eq:RSPsigma_s}\\
\sigma_\mathrm{a}^2 &= \frac{1}{4}\left( -\langle\hat{b}^2 \rangle_2 e^{2i\gamma_\mathrm{rp}}-\langle(\hat{b}^\dagger)^2 \rangle_2 e^{-2i\gamma_\mathrm{rp}}+2\langle\hat{b}^\dagger\hat{b}\rangle_2+1 \right)\, , \label{eq:RSPsigma_a}
\end{align}
where $\mathrm{arg}(\cdot)$ is the argument of a complex number and the first order moments are taken to be zero. These quantities are then fitted simultaneously to the corresponding quantities of the experimental remotely prepared states (see Fig.~\ref{fig:AntiSq} for antisqueezed variance). We are able to describe the RSP protocol presented in the main article with the parameters shown in Tab.~\ref{tab:Fit_Params}. We emphasize that the bare model only requires three fitting parameters ($n$, $r$, $n_\mathrm{f}'$) in order to obtain a good fit when estimating the remaining parameters from independent measurements. Including $\beta$, $\theta_\mathrm{f}$, $\theta_\mathrm{rp}$, and the crosstalk parameters ($\gamma^{(0)}_1$, $\kappa$, $\lambda$) as fitting parameters, only slightly improves the quantitative agreement between the experiment and the theory.

In order to derive equation~(1) in the main article, we choose $\gamma_1\,{=}\,\gamma_\mathrm{f}\,{=}\,\SI{0}{\degree}$, $\gamma_2\,{=}\,\theta_\mathrm{f}\,{=}\,\SI{90}{\degree}$, same losses after the beam splitter ($\eta_{1}\,{=}\,\eta_{2}\,{=}\,\eta$), and neglect losses before the beam splitter ($\epsilon\,{=}\,0$) as well as the effect of the electrical path lengths ($\theta_\mathrm{rp}\,{=}\,0$). Furthermore, we do not consider the experimental crosstalk ($\kappa\,{=}\,\lambda\,{=}\,0$). The protocol works optimally for fixed resources if a state with the highest purity is remotely prepared. In the limit of high JPA\,1 and JPA\,2 squeezing, $r\,{\gg}\,1$, we reach this optimal point for $G_\mathrm{f}\,{=}\,\tau/(1-\tau)$ and obtain for the optimally remotely prepared state by using equations~(\ref{eq:RSPgamma})-(\ref{eq:RSPsigma_a})
\begin{align}
\tilde{\gamma}_\mathrm{rp} &= \gamma_1 \,, \\
\tilde{\sigma}_\mathrm{s}^2 &= \frac{1}{4}\left[2(1+2n)e^{-2r}(1-\eta)\tau+2(\eta+n_\mathrm{f})\tau\right]\,, \label{eq:sigmaOptSqueezed} \\ 
\tilde{\sigma}_\mathrm{a}^2 &= \frac{(1-\eta)(1+2n)}{4 \tau\left(1+2n_\mathrm{f}\right)^2}\left\lbrace
\frac{\left(2\tau n_\mathrm{f}+1\right)^2}{2 e^{-2r}}\,{+}\,\frac{\left(2\tau n_\mathrm{f}+2\tau - 1 \right)^2}{2 e^{2 r}}\,{+}\,\frac{4 \eta  \tau ^2 n_\mathrm{f}^2+2n_\mathrm{f}\left[(1-\tau)^2+2\eta\tau^2 \right]+\eta (2\tau^2-2\tau+1)}{(1-\eta)(1+2n)}\right\rbrace\,. \label{eq:sigmaOptAnti}
\end{align}
In general, the optimal JPA\,3 gain depends on $r$ in a nontrivial manner and converges to $G_\mathrm{f}\,{=}\,\tau/(1-\tau)$ for $r\,{\to}\,\infty$. However, the latter expression offers a good approximation to the optimal JPA\,3 gain even for $r\,{\approx}\,1$ since the deviation of $\tilde{\gamma}_\mathrm{rp}$, $\tilde{\sigma}_\mathrm{s}^2$, and $\tilde{\sigma}_\mathrm{a}^2$ between the optimal JPA\,3 gain and $G_\mathrm{f}\,{=}\,\tau/(1-\tau)$ is below $1\%$ for the parameters in Tab.~\ref{tab:Fit_Params}

{\renewcommand{\arraystretch}{1.3}
\setlength{\tabcolsep}{5pt}
\begin{table*}
\caption{Model parameters used to theoretically describe the RSP protocol in the main article. The losses $\epsilon$, $\eta_1$ and $\eta_2$ are estimated from the individual loss of the components. $\gamma_2$ is fixed to the experimentally chosen squeezing angle of JPA\,2.}
\centering
\begin{ruledtabular}
\begin{tabular}{l*{13}{c}r}
$n$ & $r$ & $\gamma_{1}^{(0)}$ ($^\circ$) & $\gamma_{2}$ ($^\circ$) & $n'_\mathrm{f}$ & $\beta$ (dB) & $\epsilon$ (dB) & $\eta_{1}$ (dB) & $\eta_{2}$ (dB) & $\theta_\mathrm{f}$ ($^\circ$) & $\theta_\mathrm{rp}$ ($^\circ$) & $\kappa$ ($^\circ$) &  $\lambda$\\
\hline
$0.04$ & $1.20$ & $49.6$ & $135.0$ & $0.0059$ & $-14.6$ & $0.72$ & $0.35$ & $0.30$ & $136.5$ & $68.5$ & $-0.17$ & $0.02$ \\
\end{tabular}
\end{ruledtabular}
\label{tab:Fit_Params}
\end{table*}
}

\section{Phase space of preparable states}

\begin{figure}[ht]
\includegraphics[width=0.65\textwidth]{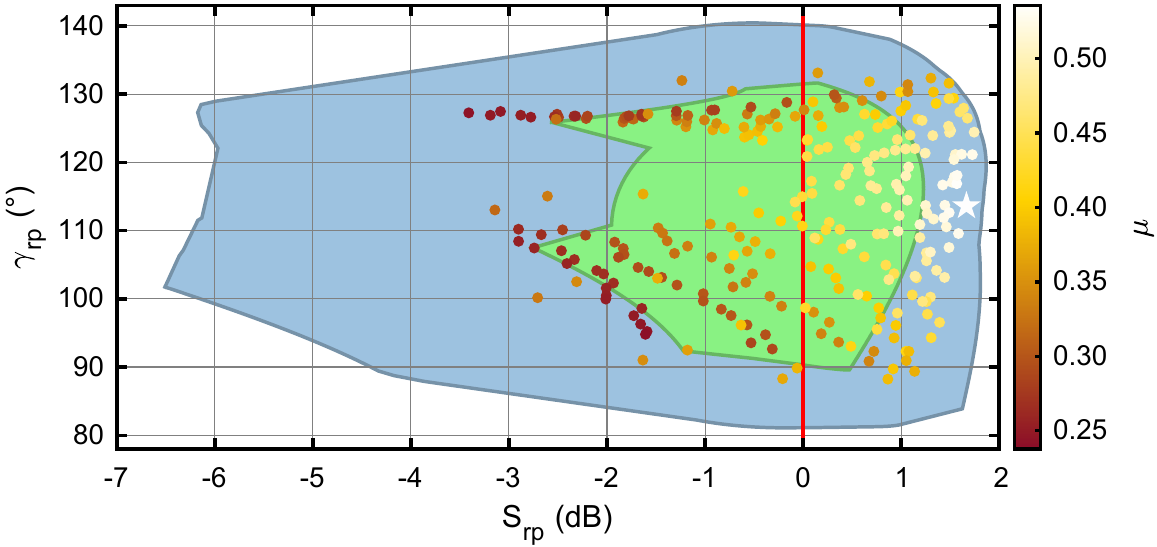}
\caption{Phase space of experimental remotely prepared states spanned by $S_\mathrm{rp}$ and $\gamma_\mathrm{rp}$. The red line marks the threshold for squeezing below the vacuum limit. The green and blue shaded area indicate the direct contour from the fit and maximum error contour, respectively. The color code indicates the purity $\mu$ of the remotely prepared states. The optimal point is marked by the white star.}
\label{fig:Contour}
\end{figure}

\noindent
The model described in the previous section allows us to theoretically investigate the phase space of the preparable states of our RSP protocol. For this purpose, we use the parameters from Tab.~\ref{tab:Fit_Params} and calculate the contour around the remotely prepared states for the experimental range of JPA\,3 gain $G_\mathrm{f}$ and amplification angle $\gamma_\mathrm{f}$ (green contour in Fig.~\ref{fig:Contour}). Alternatively, we use an iterative method to calculate the maximum error contour (blue contour in Fig.~\ref{fig:Contour}). Here, we randomly select a value from the 95\% confidence intervals of each fitting parameter and calculate the resulting contour. If the current contour lies partly or fully outside the maximum error contour, the latter is expanded so that it includes the current contour. The process is repeated until the change of the area of the maximum error contour is negligible between iterations.

We observe that the direct contour does not include all experimentally prepared states but shows a good qualitative agreement. The maximum error contour includes all measured remotely prepared states. We note that all remotely prepared states inside the contour can be continuously prepared. However, the position in the phase space does not uniformly depend on $G_\mathrm{f}$ and $\gamma_\mathrm{f}$. Since we select a finite and uniform step size of $G_\mathrm{f}$ and $\gamma_\mathrm{f}$ in the experiment, the measured remotely prepared states do not uniformly occupy the phase space.

\section{Feedforward signal}

\begin{figure}[ht]
\includegraphics[width=0.65\textwidth]{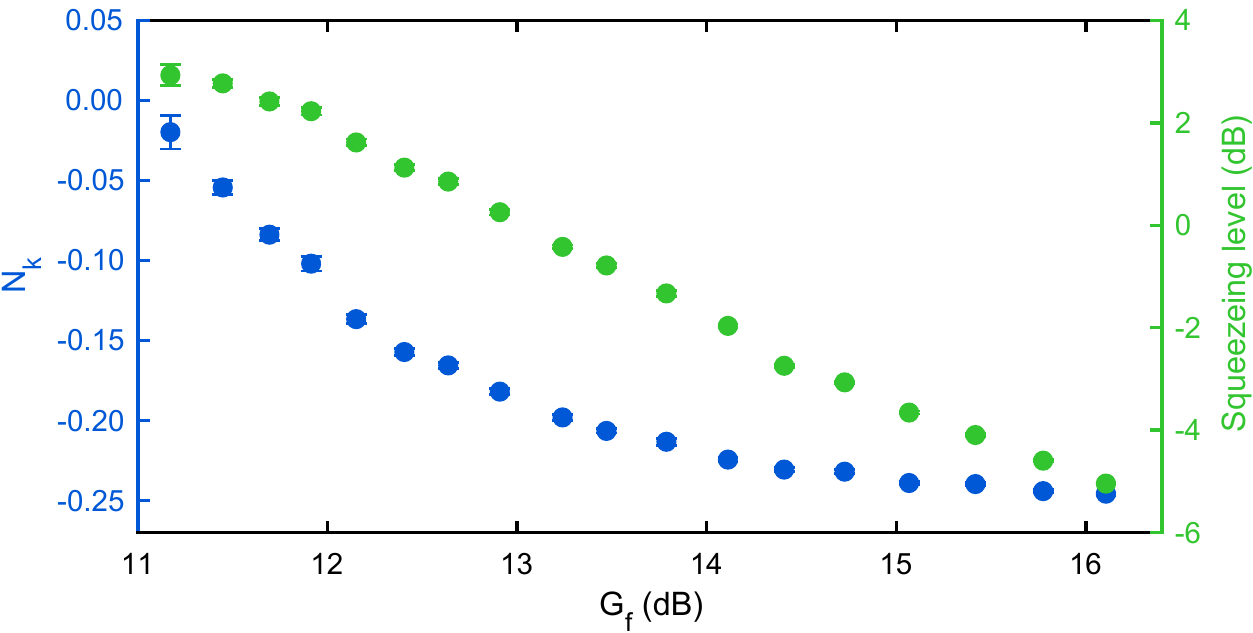}
\caption{Negativity kernel $N_\mathrm{k}$ (blue) and squeezing level (green) of the feedforward signal.}
\label{fig:FF}
\end{figure}

\noindent
The feedfoward signal is characterized by toggling the cryogenic switch into position B (see Fig.~\ref{fig:ExpSetup}) and measure both the signal from JPA\,3 (Alice's side) as well as the signal on Bob's side while all JPAs are pumped. The squeezing level of JPA\,1 and JPA\,2 for these measurements is $S\,{=}\,\SI{7.3}{\dB}$  which results in an entangled state with a negativity kernel\cite{Menzel2012,Adesso2005} $N_\mathrm{k}\,{=}\,1.8$ after the beam splitter. The JPA\,3 amplification angle is fixed to the optimal angle $\gamma_\mathrm{f}\,{=}\,\SI{0}{\degree}$. As shown in Fig.~\ref{fig:FF}, we observe no entanglement ($N_\mathrm{k}\,{\le}\,0$) between Alice and Bob after the local amplification for $G_\mathrm{f}\ge\SI{11}{dB}$. The feedforward signal is squeezed below the vacuum for low $G_\mathrm{f}$ and becomes non-squeezed above $G_\mathrm{f}\,{\simeq}\,\SI{13}{\dB}$. Our theory model and experimental evidence show that the deamplified, and possibly squeezed, quadrature has a negligible effect on the prepared state. This can be understood by considering that the feedforward signal is only weakly coupled to Bob's part of the entangled state by the directional coupler. Therefore, only the strongly amplified quadrature in the feedforward signal will affect the prepared state on Bob's side.

We consider the feedforward signal as classical if it has a positive Wigner function, is not squeezed below the vacuum and is not entangled with the signal on Bob's side. Therefore, all feedforward signals with $G_\mathrm{f}\,{\ge}\,\SI{13}{\dB}$ are classical. For $G_\mathrm{f}\,{<}\,\SI{13}{\dB}$, the information about the to-be-prepared state in the feedforward signal can be described classically as well since it is only encoded in the strongly amplified quadrature which, on its own, does not show any quantum signatures.

\section{Entropy of Gaussian states}

\noindent
The von Neumann entropy $H(X)\,{=-}\,\mathrm{Tr}(\hat{\rho}_x\mathrm{log}\,\hat{\rho}_x)$ of a quantum state $\hat{\rho}_x$ is the quantum information analogue of the entropy used in thermodynamics (up to a factor of the Boltzmann constant $k_\mathrm{B}$). For Gaussian states, $H(X)$ can be calculated from the covariance matrix $\mathbf{V}$. The von Neumann entropy of a single-mode Gaussian states is given by\cite{Serafini2004}
\begin{equation}
H(X) = f\left(\sqrt{\mathrm{det}\,\mathbf{V}}\right)\, ,
\end{equation}
where $f(x)\,{=}\,\left(2x+\frac{1}{2}\right)\mathrm{log}\left(2x+\frac{1}{2}\right)-\left(2x-\frac{1}{2}\right)\mathrm{log}\left(2x-\frac{1}{2}\right)$.

For a two-mode Gaussian state $\hat{\rho}_{AB}$, the covariance matrix can be expressed in the form
\begin{equation}
\mathbf{V}= \begin{pmatrix} \mathbf{A} & \mathbf{C}   \\ \mathbf{C}^\mathrm{T} & \mathbf{B} \end{pmatrix}\, ,
\end{equation}
where $\mathbf{A}$, $\mathbf{B}$ and $\mathbf{C}$ are $2\times2$ matrices describing the local state A, local state B and cross-correlations between both parties, respectively. From $\mathbf{V}$, one can calculate the two symplectic eigenvalues of the bipartite Gaussian state
\begin{equation}
\nu_{\pm} = \sqrt{\frac{\Delta\pm\sqrt{\Delta^2-4\mathrm{det}\,\mathbf{V}}}{2}}\, ,
\end{equation}
where $\Delta\,{=}\,\mathrm{det}\,\mathbf{A}+\mathrm{det}\,\mathbf{B}+2\mathrm{det}\,\mathbf{C}$. The entropy of the whole bipartite state is given by
\begin{equation}
H(A,B) = f\left(\nu_+\right)+f\left(\nu_-\right)\, ,
\end{equation}
and the entropy of $A$ conditioned on knowing $B$
\begin{equation}
H(A|B) = H(A,B)-H(B)
\end{equation}
is called the conditional entropy. If both parties are correlated, knowledge about $B$ will reveal information about $A$, and thus decrease its entropy, $H(A|B)\,{<}\,H(A)$.

\bibliographystyle{naturemag}
\bibliography{Bibliography_supp}